\documentclass[a4paper,10pt,twoside]{cpc-hepnp}

\usepackage{multicol}
\usepackage{graphicx}
\usepackage{booktabs}
\usepackage{amssymb,bm,mathrsfs,bbm,amscd}
\usepackage[tbtags]{amsmath}
\usepackage{lastpage}
\usepackage{CJK}

\usepackage{graphicx,color,dcolumn,booktabs,bm}
\usepackage{longtable,lscape,comment,braket}
\usepackage{txfonts}
\usepackage{overpic}
\usepackage{amssymb}
\usepackage{indentfirst}
\usepackage{feynmf}   %{feynmp}
\usepackage{slashed}  %for Feynman symbols
\usepackage{cases}
\usepackage{epstopdf}
\usepackage{psfrag}
\usepackage{subfigure}
\usepackage[colorlinks,
            citecolor=blue,
            anchorcolor=red,
            menucolor=red,
            linkcolor=red,
            filecolor=red,
            runcolor=red,
            urlcolor=blue,
            frenchlinks=red]{hyperref}

\begin{document}
\begin{CJK*}{GB}{gbsn}

\fancyhead[c]{\small Chinese Physics C~~~Vol. **, No. * (****)
******} \fancyfoot[C]{\small ******-\thepage}

\title{Exploring the two-body strong decay properties of the possible $\Lambda_cK^{*}$ and $\Sigma_cK^{(*)}$ molecules
\thanks{This work is supported by the National Natural Science Foundation of China under Grants No. 12305139 and the Xiaoxiang Scholars Programma of Hunan Normal University.}}

\author{%
      Jin-Yu Huo (»ô½ðîÚ)$^{1;1)}$\email{202220112372@hunnu.edu.cn}%
\quad Rui Chen (³ÂÈñ)$^{1,2;2)}$\email{chenrui@hunnu.edu.cn}
}
\maketitle

\address{%
$^1$Key Laboratory of Low-Dimensional Quantum Structures and Quantum Control of Ministry of Education, Department of Physics and Synergetic Innovation Center for Quantum Effects and Applications, Hunan Normal University, Changsha 410081, China\\
$^2$Hunan Research Center of the Basic Discipline for Quantum Effects and Quantum Technologies, Hunan Normal University, Changsha 410081, China
}

\begin{abstract}
In this work, we apply the effective Lagrangian approach to investigate the two-body strong decay behaviors of the possible $\Lambda_c K^*$ and $\Sigma_c K^{(*)}$ molecules, as predicted in our previous study [\href{https://doi.org/10.1103/PhysRevD.108.054011}{Phys. Rev. D \textbf{108}, 054011 (2023)}]. Our results indicate that the decay width for the coupled $\Sigma_c K / \Lambda_c K^* / \Sigma_c K^*$ molecule with $I(J^P) = 1/2(1/2^-)$ is on the order of several MeV, with the $D_s N$ channel being dominant. For the coupled $\Lambda_c K^* / \Sigma_c K^*$ molecule with $1/2(1/2^-, 3/2^-)$, the decay widths are on the order of tens of MeV, with the dominant channels being $\Sigma_c K$ and $\Sigma_c^* K$, respectively. For the $\Sigma_c K^*$ molecules with $1/2(1/2^-)$, the decay width can reach one hundred MeV, with $\Sigma_c K$ and $\Lambda_c K$ being the dominant decay channels. The decay widths for the $\Sigma_c K^*$ molecules with $1/2(3/2^-)$ and $3/2(1/2^-)$ are on the order of tens of MeV, with the dominant decay modes being $\Sigma_c^* K$ and $\Sigma_c K$, respectively. The branching ratios for all the discussed channels show little dependence on the binding energies.
\end{abstract}



\begin{multicols}{2}

\section{Introduction}\label{sec1}
{Since 2003, experiments have continuously observed numerous near-threshold $X/Y/Z/P_c/P_{cs}$ states (see review papers \cite{Chen:2016qju, Liu:2019zoy, Chen:2016spr, Guo:2017jvc, Chen:2022asf, Liu:2013waa, Hosaka:2016pey,Meng:2022ozq} for more details).} Theoretical models have proposed various explanations for these near-threshold structures, including conventional hadrons, multiquarks, hybrids, glueballs, and others. Among these explanations, the hadronic molecular scheme has attracted significant attention. As a special class of exotic states, a molecular state is composed of two or more conventional mesons and/or baryons, with its constituent hadrons bound together by strong interactions, typically resulting in a shallow binding energy. Hadronic molecules exhibit several characteristic features: their mass is usually close to the combined masses of their constituents, and they have specific quantum numbers (such as spin, parity, and isospin) that arise from the combination of their constituent hadrons. The study of hadronic molecules not only enhances our understanding of the strong interactions between conventional hadrons, but also help us explore the nature of composite particles beyond the simple quark model of hadrons.

Recently, the LHCb collaboration observed two resonances in the $D_s^+\pi^{-}(\pi^+)$ final states by performing a combined amplitude analysis on the decays $B^0\to\bar{D}^0D_s^+\pi^-$ and $B^+\to D^-D_s^+\pi^+$ \cite{LHCb:2022lzp,LHCb:2022sfr}. Both states have spin-parity of $J^P=0^+$. Their masses and widths are:
\begin{eqnarray*}\left.\begin{array}{ll}
T_{c\bar{s}}^{a0}(2900): \quad\quad\quad &M = 2892 \pm 14 \pm 15 \text{MeV}, \\ &\Gamma = 119 \pm 26 \pm 12 \text{MeV},\\
T_{c\bar{s}}^{a++}(2900): &M = 2921 \pm 17 \pm 19 \text{MeV}, \\
   &\Gamma = 137 \pm 32 \pm 14 \text{MeV}. \end{array}\right.
\end{eqnarray*}
According to their mass positions, quantum numbers, and  the decay channels, $T_{c\bar{s}}^{a0}(2900)$ and $T_{c\bar{s}}^{a++}(2900)$ belong to the same isovector triplet, and the simplest valance quark components are $c\bar{s}d\bar{u}$ and $c\bar{s}\bar{d}u$, respectively. In addition to $T_{c\bar{s}}^{a0}(2900)$ and $T_{c\bar{s}}^{a++}(2900)$, two other interesting charm-strange structures were observed in the zai decay channel: $D_{s0}(2317)$ \cite{BaBar:2003oey, Belle:2003kup, BaBar:2006eep} and $D_{s1}(2460)$ \cite{CLEO:2003ggt, Belle:2003kup, BaBar:2006eep, BaBar:2003cdx}. It is worth to notice that these states are very close to the mass thresholds of the charmed meson and strange meson, which has inspired the proposal of hadronic molecular explanations for them (see Refs. \cite{Chen:2016spr, Guo:2017jvc} for a detailed review of the different assignments).

Exploring the molecular partners can be one of important way to verify the hadronic molecular interpretations of these states. Once replacing the light quark in the charmed meson by light diquark, one can extend to search for the charm-strange pentaquark $P_{c\bar{s}}$ partners composed of $Y_cK^{(*)} (Y_c=\Lambda_c, \Sigma_c)$ systems. In our previous work \cite{Chen:2023qlx}, we studied the $Y_cK^{(*)}$ interactions using the one-boson-exchange model, considering both $S-D$ wave mixing effects and coupled-channel effects, and we can predict the existences of possible charm-strange molecular pentaquarks, including the single $\Sigma_cK^*$ molecular states with $I(J^P)=1/2(1/2^-)$, $1/2(3/2^-)$ and $3/2(1/2^-)$, the coupled $\Lambda_cK^*/\Sigma_cK^*$ molecular states with $1/2(1/2^-)$ and $1/2(3/2^-)$, and the coupled $\Sigma_cK/\Lambda_cK^*/\Sigma_cK^*$ molecular state with $1/2(1/2^-)$. When we solved the coupled channel Schr\"{o}dinger equations to explore the loosely bound states mainly composed of the higher channels, like the $\Lambda_cK^*$ and $\Sigma_cK^*$, the lower channels were not included, which have relatively weak coupling compared to the dominant channels responsible for forming the state, generally.

A comprehensive study of the properties of these molecular states will provide more valuable information, which can offer guidance for future experimental investigations. In this work, we further study the two-body strong decay behaviors for the predicted $P_{c\bar{s}}$ molecules by employing the effective Largrangians approach. Since the coupled-channel effects play a crucial role in binding the coupled $P_{c\bar{s}}$ molecular candidates, we also consider these effects in our analysis.

As is well known, the decay modes of a hadronic molecule reflect the possible ways in which the constituent hadrons can interact or decay. Theorists have proposed different models to explore the decay properties for the possible molecules, such as the effective Lagrangian approach \cite{Chen:2021tip,Chen:2017xat,Meng:2021jnw,Wang:2024kke,Lin:2019qiv,Shen:2019evi,Lin:2017mtz,Xiao:2019mvs,Guo:2019fdo,Dong:2009tg,Huang:2018wgr,Yue:2024paz,Yue:2022gym,Chen:2017abq,Liu:2024ugt,Ling:2021bir,Xie:2022hhv,Li:2023zag}, the constituent quark model \cite{Sheng:2024hkf,Wang:2022nqs}, the quark interchange model \cite{Wang:2019spc,Liu:2014eka,Yang:2021sue,Zhou:2019swr,Wang:2018pwi}, the spin rearrangement scheme in the heavy quark limi t\cite{Ma:2014zva,Ma:2014ofa}, the heavy quark spin symmetry \cite{Sakai:2019qph,Voloshin:2019aut,Burns:2021jlu}, the
 QCD sum rule \cite{Xu:2019zme,Wang:2023ews}, Fierz rearrangement \cite{Chen:2020opr,Chen:2020pac} and the unitary approach \cite{Yang:2024nss}. We hope that our investigation of the decay properties will help confirm the existence and identify the nature of these predicted molecules.

This paper is organized as follows. After this introduction, we present the two-body strong decay amplitudes for the predicted charm-strange $Y_cK^{(*)}$ molecular pentaquarks in Sec.~\ref{sec2}. In Sec.~\ref{sec3}, we provide the corresponding numerical results. The paper ends with a summary in Sec. \ref{sec4}.

\section{Formalism}\label{sec2}
\begin{center}
% Requires \usepackage{graphicx}
  %\includegraphics[width=5.4in]{XisD1.eps}
\includegraphics[width=3.0in]{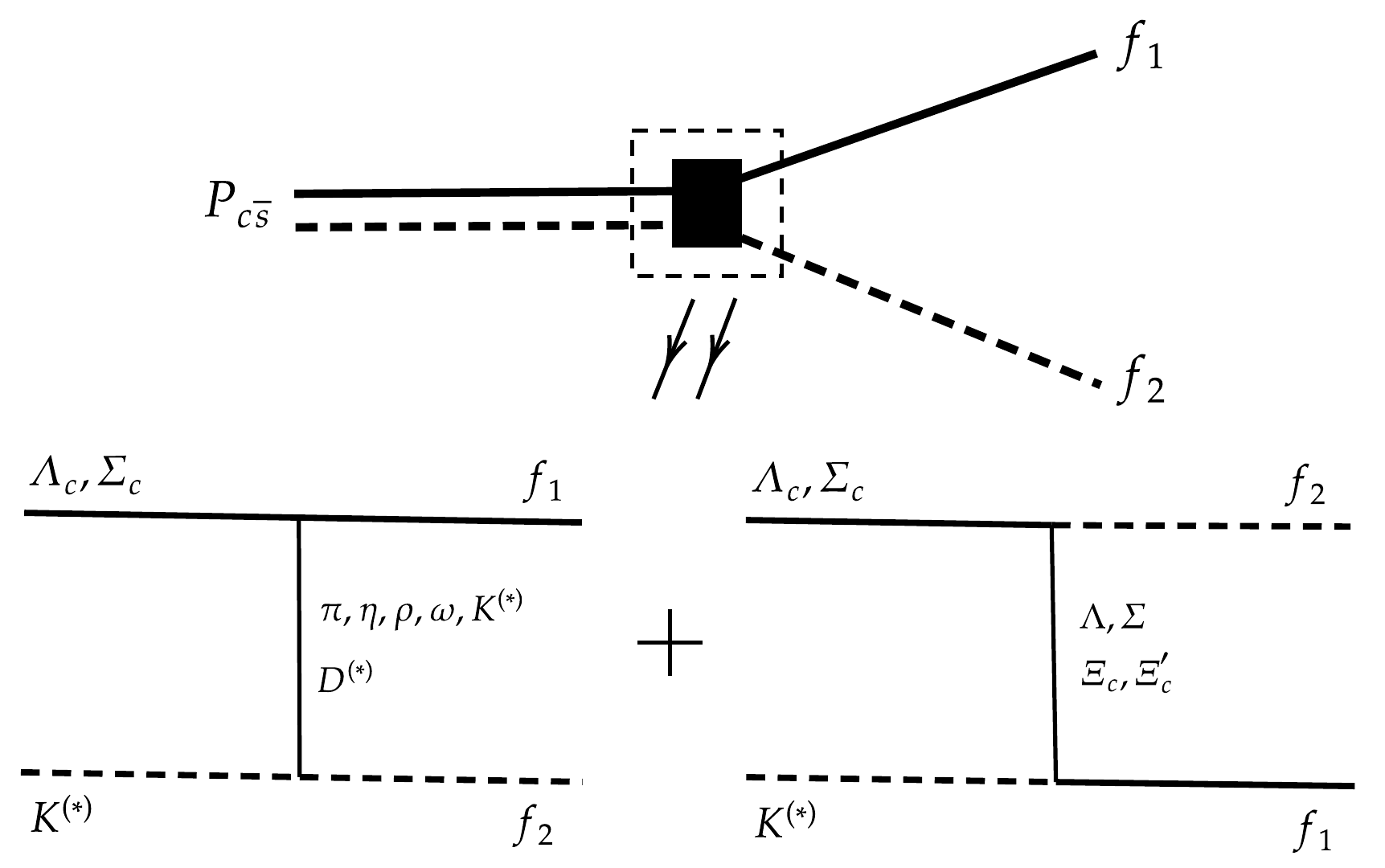}
\figcaption{\label{diagram}The diagram for the $P_{c\bar{s}}$ molecules decaying into $f_{1}$ and $f_{2}$ final states.}
\end{center}
In this work, we focus on the two-body strong decay properties for the predicted $P_{c\bar{s}}$ molecules via $S-$wave interactions. In Figure \ref{diagram}, we present the relevant diagram, where the interactions can occur through the exchange of either a meson (left side) or a baryon (right side) between the molecular constituents.

For a hadronic molecule composed of several channels $|i\rangle=\sum_n|A_nB_n\rangle$, where $A_n$ and $B_n$ denote the molecular constitutions in $n-$th channel, the two-body decay interactions for the $i\to f_1+f_2$ process are related to the interactions for the $A_n+B_n\to f_1+f_2$ process, as follows:
\begin{eqnarray}
\langle f_1f_2|V|i\rangle &=&\sum_n\langle f_1f_2|V|A_nB_n\rangle\langle A_nB_n|i\rangle\nonumber\\
 &=&\sum_n\int\frac{d^3k d^3r}{(2\pi)^3}e^{-i\bm{k}\cdot\bm{r}}
 \psi_{A_nB_n}(\bm{r})\langle f_1f_2|V|A_nB_n\rangle.\,\,
\end{eqnarray}
Here, $\psi_{A_nB_n}(\bm{r})$ is the wave functions for the $n-$th channels in the $r-$coordinate space. Considering the different normalization conventions used for the scattering amplitude, one have
\begin{eqnarray}
&&\langle f_1f_2|V|i\rangle = -\frac{\mathcal{M}(i\to f_1+f_2)}{\sqrt{2E_{i}}\sqrt{2E_{f_1}}\sqrt{2E_{f_2}}},\\
&&\langle f_1f_2|V|A_nB_n\rangle = \nonumber\\
&&\quad\quad\quad\,-\frac{\mathcal{M}\left(A_n(\bm{k})+B_n(-\bm{k})\to f_1(\bm{p})+f_2(-\bm{p})\right)}{\sqrt{2E_{A_n}}
\sqrt{2E_{B_n}}\sqrt{2E_{f_1}}\sqrt{2E_{f_2}}}.\quad\,\,
\end{eqnarray}

In the rest frame of the initial state, the partial decay width for the $i\to f_1+f_2$ process can be expressed as
\begin{eqnarray}
d\Gamma &=& \frac{1}{2J+1}\frac{|\bm{p}|}{32\pi^2 m_{i}^2}|\mathcal{M}(i\to f_1+f_2)|^2d\Omega,
\end{eqnarray}
where $J$ and $m_i$ stand for the spin and mass for the initial state, respectively. $\bm{p}$ is the three momentum for the final states, which reads as
\begin{eqnarray*}
|\bm{p}| &=& \sqrt{\left(m_{i}^2-(m_{f_1}+m_{f_2})^2\right)\left(m_{i}^2-(m_{f_1}-m_{f_2})^2\right)}/(2m_{i}).
\end{eqnarray*}
And $\mathcal{M}(i\to f_1+f_2)$ is the scattering amplitude. In Table \ref{channels}, we collect the corresponding amplitudes for the molecular constitutions decaying into two-body final states via the $S-$wave interactions.

The relevant effective Lagrangians can be constructed as \cite{Wang:2022oof}
\begin{eqnarray}
\mathcal{L}_{PPV} &=& i\sqrt{2}g_{PPV}(P\partial^{\mu}{P}-\partial^{\mu}{P}{P}){V}_{\mu},\\
\mathcal{L}_{VVP} &=&\frac{g_{VVP}}{m_V} \epsilon_{\mu\nu\alpha\beta}
    \partial^{\mu}{V}^{\nu}\partial^{\alpha}{V}^{\beta}{P},\\
\mathcal{L}_{VVV} &=& ig_{VVV}
    \langle{V}^{\mu}[{V}^{\nu},\partial_{\mu}V_\nu)]\rangle,\\
\mathcal{L}_{BBP} &=& \frac{g_{BBP}}{m_P}\bar{{B}}\gamma^\mu\gamma^5\partial_\mu {P}{B},\label{lag1}\\
\mathcal{L}_{BBV} &=& -g_{BBV}\bar{{B}}(\gamma^{\mu}
              -\frac{k}{2m_{B}}\sigma^{\mu\nu}\partial_{\nu}){V}_{\mu}{B},\label{lag2}\\
\mathcal{L}_{BDP} &=&- \frac{g_{BDP}}{m_{{P}}}
             (\bar{{B}}\partial^{\mu}{P}{{D}}_{\mu}+\bar{D}_{\mu}\partial^{\mu}{P}B),\\\nonumber\\
\mathcal{L}_{BDV} &=& i\frac{g_{BDV}}{m_{{V}}}
        [\bar{{B}}\gamma_{\mu}\gamma^5{{D}}_{\nu}(\partial^{\mu}{V}^{\nu}-\partial^{\nu}{V}^{\mu}) \nonumber\\
        &&+\bar{{D}}_{\mu}\gamma_{\nu}\gamma^5{B}(\partial^{\mu}{V}^{\nu}-\partial^{\nu}{V}^{\mu})].\label{lag2}
\end{eqnarray}

\end{multicols}

\begin{center}
\tabcaption{\label{channels} The amplitudes for the $Y_cK^{(*)}$ decaying into two-body final states via the $S-$wave interactions.}
\footnotesize
\renewcommand{\arraystretch}{2}
\begin{tabular*}{177mm}{@{\extracolsep{\fill}}ccllllllll}
\toprule[1pt]\toprule[1pt]
  Initial state  &Final state  &Amplitudes
 \\\hline
$\Sigma_cK  1/2(1/2^-)$    &$D_sN$               &$-\sqrt{3}\mathcal{M}_{BP\to BP}^{D^*}-\sqrt{3}\mathcal{M}_{BP\to PB}^{\Sigma}$ \\
                               &$\Lambda_cK$                &$-\frac{3}{\sqrt{6}}\mathcal{M}_{BP\to BP}^{\rho}-\sqrt{3}\mathcal{M}_{BP\to PB}^{\Xi_c}-\sqrt{3}\mathcal{M}_{BP\to PB}^{\Xi_c^\prime}$                \\
$\Lambda_c K^*1/2(1/2^-)$ &$D_sN$               &$\mathcal{M}_{BV\to BP}^{D}+\mathcal{M}_{BV\to BP}^{D^*}+\mathcal{M}_{BV\to PB}^{\Lambda}$            \\
                                 &$D_s^*N$             &$\mathcal{M}_{BV\to BV}^{D}+\mathcal{M}_{BV\to BV}^{D^*}+\mathcal{M}_{BV\to VB}^{\Lambda}$              \\   \

                                  &$\Lambda_cK$             &$\frac{3}{\sqrt{6}}\mathcal{M}_{BV\to BP}^{\eta}-\frac{1}{\sqrt{2}}\mathcal{M}_{BV\to BP}^{\omega}+\mathcal{M}_{BV\to PB}^{\Xi_c}+\mathcal{M}_{BV\to PB}^{\Xi_c^\prime}$ \\
                              &$\Sigma_cK$                &$-\frac{3}{\sqrt{6}}\mathcal{M}_{BV\to BP}^{\pi}-\frac{3}{\sqrt{6}}\mathcal{M}_{BV\to BP}^{\rho}-\sqrt{3}\mathcal{M}_{BV\to PB}^{\Xi_c}-\sqrt{3}\mathcal{M}_{BV\to PB}^{\Xi_c^\prime}$\\
$\Lambda_c K^*1/2(3/2^-)$ &$D_s^*N$    &$\mathcal{M}_{BV\to BV}^{D}+\mathcal{M}_{BV\to BV}^{D^*}+\mathcal{M}_{BV\to VB}^{\Lambda}$\\
                            &$\Sigma_c^*K$     &$-\frac{3}{\sqrt{6}}\mathcal{M}_{BV\to DP}^{\pi}-\frac{3}{\sqrt{6}}\mathcal{M}_{BV\to DP}^{\rho}-\sqrt{3}\mathcal{M}_{BV\to PD}^{\Xi_c}-\sqrt{3}\mathcal{M}_{BV\to PD}^{\Xi_c^\prime}$\\
$\Sigma_c K^*1/2(1/2^-)$ &$D_sN$     &$-\sqrt{3}\mathcal{M}_{BV\to BP}^{D}-\sqrt{3}\mathcal{M}_{BV\to BP}^{D^*}-\sqrt{3}\mathcal{M}_{BV\to PB}^{\Sigma}$\\
                                  &$D_s^*N$      &$-\sqrt{3}\mathcal{M}_{BV\to BV}^{D}-\sqrt{3}\mathcal{M}_{BV\to BV}^{D^*}-\sqrt{3}\mathcal{M}_{BV\to VB}^{\Sigma}$\\
                                &$\Lambda_cK$      &$-\frac{3}{\sqrt{6}}\mathcal{M}_{BV\to BP}^{\pi}-\frac{3}{\sqrt{6}}\mathcal{M}_{BV\to BP}^{\rho}-\sqrt{3}\mathcal{M}_{BV\to PB}^{\Xi_c}-\sqrt{3}\mathcal{M}_{BV\to PB}^{\Xi_c^\prime}$\\
                                 &$\Lambda_cK^*$    &$-\frac{3}{\sqrt{6}}\mathcal{M}_{BV\to BV}^{\pi}-\frac{3}{\sqrt{6}}\mathcal{M}_{BV\to BV}^{\rho}-\sqrt{3}\mathcal{M}_{BV\to VB}^{\Xi_c}-\sqrt{3}\mathcal{M}_{BV\to VB}^{\Xi_c^\prime}$\\
                             &$\Sigma_cK$     &$-\sqrt{2}\mathcal{M}_{BV\to BP}^{\pi}+\frac{3}{\sqrt{6}}\mathcal{M}_{BV\to BP}^{\eta}-\sqrt{2}\mathcal{M}_{BV\to BP}^{\rho}-\frac{1}{\sqrt{2}}\mathcal{M}_{BV\to BP}^{\omega}-\mathcal{M}_{BV\to PB}^{\Xi_c}-\mathcal{M}_{BV\to PB}^{\Xi_c^\prime}$\\
$\Sigma_c K^*1/2(3/2^-)$    &$D_s^*N$      &$-\sqrt{3}\mathcal{M}_{BV\to BV}^{D}-\sqrt{3}\mathcal{M}_{BV\to BV}^{D^*}-\sqrt{3}\mathcal{M}_{BV\to VB}^{\Sigma}$\\
                               &$\Lambda_cK^*$    &$-\frac{3}{\sqrt{6}}\mathcal{M}_{BV\to BV}^{\pi}-\frac{3}{\sqrt{6}}\mathcal{M}_{BV\to BV}^{\rho}-\sqrt{3}\mathcal{M}_{BV\to VB}^{\Xi_c}-\sqrt{3}\mathcal{M}_{BV\to VB}^{\Xi_c^\prime}$\\
                            &$\Sigma_c^*K$     &$-\sqrt{2}\mathcal{M}_{BV\to DP}^{\pi}+\frac{3}{\sqrt{6}}\mathcal{M}_{BV\to DP}^{\eta}-\sqrt{2}\mathcal{M}_{BV\to DP}^{\rho}-\frac{1}{\sqrt{2}}\mathcal{M}_{BV\to DP}^{\omega}-\mathcal{M}_{BV\to PD}^{\Xi_c}-\mathcal{M}_{BV\to PD}^{\Xi_c^\prime}$\\
 $\Sigma_c K^*3/2(1/2^-)$    &$\Sigma_cK$      &$\frac{1}{\sqrt{2}}\mathcal{M}_{BV\to BP}^{\pi}+\frac{3}{\sqrt{6}}\mathcal{M}_{BV\to BP}^{\eta}+\frac{1}{\sqrt{2}}\mathcal{M}_{BV\to BP}^{\rho}-\frac{1}{\sqrt{2}}\mathcal{M}_{BV\to BP}^{\omega}+2\mathcal{M}_{BV\to PB}^{\Xi_c}+2\mathcal{M}_{BV\to PB}^{\Xi_c^\prime}$\\
\bottomrule[1pt]\bottomrule[1pt]
\end{tabular*}%
\end{center}

\vspace{4mm}

\begin{center}
\tabcaption{\label{coupling} The coupling constants adopted in our calculations. Here, we take $g_{NN\rho}=3.25$, $f_{NN\rho}=19.82$, $g_{NN\pi}=0.989$, $g_{\Delta N\pi}=2.13$, $g_{\Delta N\rho}=16.03$, $g_{VVP}=-7.07$, $g_{PPV}=3.02$, $g_{VVV}=2.30 $, and $\alpha_{BBV}=1.15$ \cite{Wang:2022oof}.}
\footnotesize
\renewcommand{\arraystretch}{2}
\begin{tabular*}{177mm}{@{\extracolsep{\fill}}lllll}
\toprule[1pt]\toprule[1pt]
$g_{\Lambda_c\Sigma_c^*\pi}=\frac{1}{\sqrt{2}}g_{\Delta N\pi}$     &$g_{\Sigma NK}=\frac{1}{5}g_{NN\pi}$                                       &$g_{\Sigma_c\Sigma_c\eta}=\frac{4}{5\sqrt{3}}g_{NN\pi}$   &$f_{\Sigma_c\Lambda_c\rho}=\frac{\sqrt{3}}{2}f_{NN\rho}$ &$g_{\Sigma_c\Lambda_c\rho}=\frac{2}{\sqrt{3}}(1-\alpha_{BBV})g_{NN\rho}$  \\
 $g_{\Sigma_c\Sigma^*_c\pi}=\frac{1}{\sqrt{6}}g_{\Delta N\pi}$                                  &$g_{\Lambda_c N D}=-\frac{3\sqrt{3}}{5}g_{NN\pi}$   &$g_{\Sigma_c\Sigma_c\pi}=\frac{4}{5}g_{NN\pi}$     &$f_{\Sigma_cND^*}=\frac{1}{2}f_{NN\rho}$ &$g_{\Sigma_c\Sigma D_s^*}=\sqrt{2}(2\alpha_{BBV}-1)g_{NN\rho}$\\
$g_{\Sigma^*_c\Xi_cK^*}=-\frac{1}{2}g_{\Delta N\rho}$ &$g_{\Lambda_c \Lambda D_s}=\frac{3\sqrt{2}}{5}g_{NN\pi}$                      &$g_{\Sigma_c\Sigma_c\rho}=2\alpha_{BBV}g_{NN\rho}$    &$f_{\Sigma N K^*}=\frac{1}{2}f_{NN\rho}$   &$g_{\Sigma N K^*}=(1-2\alpha_{BBV})g_{NN\rho}$      \\
$g_{\Sigma^*_c\Xi_c^{\prime}K^*}=\frac{1}{2\sqrt{3}}g_{\Delta N\rho}$     &$g_{\Sigma_c\Sigma D_s}=-\frac{\sqrt{2}}{5}g_{NN\pi}$   & $g_{\Sigma_cDN}=\frac{1}{5}g_{NN\pi}$      & $f_{\Lambda N K^*}=-\frac{\sqrt{3}}{2}f_{NN\rho}$    &$g_{\Lambda N K^*}=-\frac{1}{\sqrt{3}}(1+2\alpha_{BBV})g_{NN\rho}$\\
$g_{\Sigma^*_c\Xi_cK}=-\frac{1}{2}g_{\Delta N\pi}$ &$f_{\Lambda_c ND^*}=-\frac{\sqrt{3}}{2}f_{NN\rho}$                              &$g_{\Sigma_c\Sigma^*_c\eta}=\frac{1}{2\sqrt{2}}g_{\Delta N\pi}$      &$f_{\Lambda_c\Lambda D_s^*}=\frac{1}{\sqrt{2}}f_{NN\rho}$  &$g_{\Lambda_c \Lambda D_s^*}=\frac{\sqrt{2}}{3}(1+2\alpha_{BBV}{})g_{NN\rho}$\\
  $g_{\Sigma_c^*\Xi_c^{\prime}K}=\frac{1}{2\sqrt{3}}g_{\Delta N\pi}$             &$g_{\Sigma_c\Sigma^*_c\omega}=-\frac{1}{\sqrt{6}}g_{\Delta N\rho}$
&$g_{\Lambda_c\Sigma_c\pi}=\frac{2\sqrt{3}}{5}g_{NN\pi}$    &$f_{\Sigma_c\Sigma_c\omega}=\frac{1}{2}f_{NN\rho}$      &$g_{\Lambda_c\Lambda_c\omega}=\frac{2}{3}(5\alpha_{BBV}-2)g_{NN\rho}$ \\
  $g_{\Sigma_c\Sigma_c\omega}=2\alpha_{BBV}g_{NN\rho}$     &$g_{\Lambda_c\Xi_c^{\prime}K}=\frac{\sqrt{6}}{5}g_{NN\pi}$           &$g_{\Lambda_c\Sigma_c^*\rho}=\frac{1}{\sqrt{2}}g_{\Delta N\rho}$   &$f_{\Lambda_c\Lambda_c\omega}=-\frac{1}{2}f_{NN\rho}$  &  $g_{\Lambda_c N D^*}=-\frac{1}{\sqrt{3}}(1+2\alpha_{BBV})g_{NN\rho}$\\

 $f_{\Sigma_c\Xi_c^{\prime}K^*}=\frac{1}{2\sqrt{2}}f_{NN\rho}$      &  $f_{\Lambda_c\Xi_cK^*}=-\frac{1}{2\sqrt{2}}f_{NN\rho}$  &$f_{\Sigma_c\Sigma_c\rho}=\frac{1}{2}f_{NN\rho}$  & $f_{\Sigma_c\Sigma D_s^*}=-\frac{1}{\sqrt{2}}f_{NN\rho}$   &$g_{\Lambda_c\Xi_cK^*}=\frac{\sqrt{2}}{3}(5\alpha_{BBV}-2)g_{NN\rho}$  \\
$g_{\Sigma_c\Xi_cK}=\frac{\sqrt{6}}{5}g_{NN\pi}$           & $g_{\Sigma_c\Xi_c^{\prime}K}=\frac{2\sqrt{2}}{5}g_{NN\pi}$  &$f_{\Lambda_c\Xi_c^{\prime}K^*}=\frac{1}{2}\sqrt{\frac{3}{2}}f_{NN\rho}$  &$g_{\Sigma_c\Xi_c^{\prime}K^*}=\sqrt{2}\alpha_{BBV}g_{NN\rho}$  &$g_{\Lambda_c\Xi_c^{\prime}K^*}=\sqrt{\frac{2}{3}}(1-\alpha_{BBV})g_{NN\rho}$  \\
        $g_{\Sigma_c\Sigma^*_c\rho}=\frac{1}{\sqrt{6}}g_{\Delta N\rho}$   &$g_{\Lambda_c \Lambda D_s}=\frac{3\sqrt{2}}{5}g_{NN\pi}$   &$f_{\Sigma_c\Xi_cK^*}=\frac{1}{2}\sqrt{\frac{3}{2}}f_{NN\rho}$        & $g_{\Sigma_cND^*}=(1-2\alpha_{BBV})g_{NN\rho}$&$g_{\Sigma_c\Xi_cK^*}=\sqrt{\frac{2}{3}}(1-\alpha_{BBV})g_{NN\rho}$\\
\bottomrule[1pt]\bottomrule[1pt]
\end{tabular*}%
\end{center}

\begin{multicols}{2}
\noindent Here, ${P}$, ${V}$, ${B}$, and ${D}$ stand for the pseudoscalar and vector mesons, octet, and decuplet baryons. The coupling constants are estimated by the SU(4) flavor symmetry. In Table \ref{coupling}, we collect the values of all the coupling constants.

According to the above effective Lagrangians, one can write down the scattering amplitudes for the molecular constitutions decaying to the two-body final states by exchanging one particle as follows:
\begin{eqnarray}
\mathcal{M}_{BP\to BP}^{V} &=& \left\{g_{BBV}\bar{u}_3\gamma^{\mu}u_1
      +\frac{f_{BBV}}{4m_B}\bar{u}_3(\gamma^{\mu}\gamma^{\nu}-\gamma^{\nu}\gamma^{\mu})q_{\nu}u_1\right\}\nonumber\\
     &&\times\frac{g_{\mu\beta}-q_{\mu}q_{\beta}/m_{V}^2}{q^2-m_{V}^2}\left\{-g_{PPV}(p_4^{\beta}+p_2^{\beta})\right\},\\ \nonumber
\end{eqnarray}
\begin{eqnarray}
\mathcal{M}_{BP\to BP}^{B} &=& \frac{g_{BBP}}{m_P}\bar{u}_3\gamma^\alpha p_{1\alpha}\gamma_5\frac{1}{\rlap\slash{q}-m_{B}}\frac{g^\prime_{BBP}}{m^\prime_P}\gamma^\mu p_{4\mu}\gamma_5u_2,\\ \nonumber\\
\mathcal{M}_{{BP\to BV}}^{{P}} &=& \frac{g_{BBP}}{m_P}\bar{u}_3\gamma_5 \rlap\slash{q} u_1\frac{{1}}{q^2-m_{B}^2}g_{PPV}\epsilon_4^{\mu\dag}(q_{\mu}-p_{2\mu})
      ,\\ \nonumber\\
\mathcal{M}_{{BP\to BV}}^{{V}} &=& \left\{g_{BBV}\bar{u}_3\gamma^{\mu}u_1+\frac{f_{BBV}}{4m_{B}}\bar{u}_3(\gamma^{\mu}\gamma^{\nu}
      -\gamma^{\nu}\gamma^{\mu})q_{\nu}u_1\right\}\nonumber\\
      &&\times\frac{g_{\mu\beta}-q_{\mu}q_{\beta}/m_{V}^2}{q^2-m_{V}^2}
      \frac{g_{VVP}}{m_V}\varepsilon^{\lambda\nu\alpha\beta}p_{4\nu}\epsilon_{4\lambda}^{\dag}q_{\alpha},\\\nonumber\\
%%%
\mathcal{M}_{{BP\to VB}}^{{B}} &=& \frac{g_{BBP}}{m_P}\bar{u}_4\gamma^\mu p_{2\mu}\gamma_5\rlap\slash{q}\frac{1}{\rlap\slash{q}-m_{B}}
     \left\{g_{BBV}\epsilon_{3\mu}^{\dag}\gamma^{\mu}u_1\right.\nonumber\\
     &&\left.-\frac{f_{BBV}}{4m_{B}}p_{3\mu}\epsilon_{3\nu}^{\dag}
      (\gamma^{\mu}\gamma^{\nu}-\gamma^{\nu}\gamma^{\mu})q_{\nu}u_1\right\},\\\nonumber\\
%%%
%\mathcal{M}_{\mathbb{BV\to BP}}^{\mathbb{P}} &=& -g_p\bar{u}_3\gamma_5u_1\frac{1}{q^2-m_{\mathbb{P}}^2}g_{PPV}\epsilon_2^{\mu}(p_{4\mu}+q_{\mu}),\\
%%%
%\mathcal{M}_{\mathbb{BV\to BP}}^{\mathbb{V}} &=& \left\{g_v\bar{u}_3\gamma^{\mu}u_1
%      +\frac{f_v}{4m^*}\bar{u}_3(\gamma^{\mu}\gamma^{\nu}-\gamma^{\nu}\gamma^{\mu})q_{\nu}u_1\right\}\nonumber\\
%      &&\times\frac{g_{\mu\beta}-q_{\mu}q_{\beta}/m_{\mathbb{V}}^2}{q^2-m_{\mathbb{V}}^2}
%      g_{VVP}\varepsilon^{\lambda\sigma\alpha\beta}p_{2\lambda}\epsilon_{2\sigma}p_{4\alpha},\,\,\\
%%%
%\mathcal{M}_{\mathbb{BV\to BP}}^{\mathbb{B}} &=& \left\{g_v\bar{u}_4\gamma^{\mu}\epsilon_{2\mu}
%      +\frac{f_v}{4m^*}\bar{u}_4(\gamma^{\mu}\gamma^{\nu}-\gamma^{\nu}\gamma^{\mu})p_{2\mu}\epsilon_{2\nu}\right\}\nonumber\\
%      &&\times\frac{-1}{\rlap\slash{q}-m_{\mathbb{B}}}g_p\gamma_5u_1,\\
%%%
\mathcal{M}_{{BV\to BV}}^{{P}} &=& -\frac{g_{BBP}}{m_P }\bar{u}_3\gamma_5\gamma^\mu q_\mu u_1\frac{1}{q^2-m_{P}^2}\frac{g_{VVP}}{m_V}\varepsilon^{\lambda\sigma\alpha\beta}\nonumber\\\
&&\times p_{4\lambda}\epsilon_{4\sigma}^{\dag}p_{2\alpha}\epsilon_{2\beta}
,\\\nonumber\\
%%%
\mathcal{M}_{{BV\to BV}}^{{V}} &=& \left\{g_{BBV}\bar{u}_3\gamma^{\mu}u_1
      +\frac{f_{BBV}}{4m_B}\bar{u}_3(\gamma^{\mu}\gamma^{\nu}-\gamma^{\nu}\gamma^{\mu})q_{\nu}u_1\right\}\nonumber\\
      &&\times\frac{g_{\mu\beta}-q_{\mu}q_{\beta}/m_{V}^2}{q^2-m_{V}^2}
      g_{VVV}\left\{\epsilon_{4}^{\alpha\dag}\epsilon_{2}^{\beta}(p_{2\alpha}-q_{\alpha})\right.\nonumber\\
      &&\left.
      -\epsilon_{2\alpha}\epsilon_4^{\alpha\dag}(p_2^{\beta}+p_4^{\beta})
      +\epsilon_{2\alpha}(\epsilon_4^{\beta\dag}q^{\alpha}+p_4^{\alpha}\epsilon_4^{\beta\dag})\right\},\\\nonumber\\
%%%
\mathcal{M}_{{BV\to VB}}^{{B}} &=& \left\{g_{BBV}\bar{u}_4\gamma^{\mu}\epsilon_{2\mu}
      +\frac{f_{BBV}}{4m_B}\bar{u}_4(\gamma^{\mu}\gamma^{\nu}-\gamma^{\nu}\gamma^{\mu})p_{2\mu}\epsilon_{2\nu}\right\}\nonumber\\
      &&\times\frac{1}{\rlap\slash{q}-m_{B}}
      \left\{g^\prime_{BBV}\gamma^{\alpha}\epsilon_{3\alpha}^{\dag}u_1\right.\nonumber\\
      &&\left.+\frac{f_{BBV}'}{4m'}(\gamma^{\alpha}\gamma^{\beta}-\gamma^{\alpha}\gamma^{\beta})p_{3\alpha}\epsilon_{3\beta}^{\dag}u_1\right\},\\\nonumber\\
%%%
%\mathcal{M}_{\mathbb{DP\to BP}}^{\mathbb{V}} &=& \frac{g_{BDV}}{m_V}\bar{u}_3\gamma^5(\gamma^{\nu}u_{1}^{\mu}-\gamma^{\mu}u_1^{\nu})q_{\mu}\nonumber\\
%      &&\times\frac{g_{\nu\beta}-q_{\nu}q_{\beta}/m_{\mathbb{V}}^2}{q^2-m_{\mathbb{V}}^2}
%      g_{PPV}(p_4^{\beta}+p_2^{\beta}),\\
%%%
%\mathcal{M}_{\mathbb{DP\to BP}}^{\mathbb{B}} &=& -g_{BBP}\bar{u}_4\gamma^5\frac{1}{\rlap\slash{q}-m_{\mathbb{B}}}
%      \frac{g_{BDP}}{m_P}q_{\mu}u_1^{\mu},\\
%%%
\mathcal{M}_{{DP\to VB}}^{{B}} &=& i\frac{g_{BBP}}{m_P}\bar{u}_4\gamma^5 \gamma_\alpha p_{2}^\alpha \nonumber\\
&&\times \frac{1}{\rlap\slash{q}-m_{B}} \frac{g_{BDV}}{m_V} \gamma^5(\gamma^{\nu}u_{1}^{\mu}-\gamma^{\mu}u_1^{\nu})q_{\mu}\epsilon_{3\nu}^{\dag},\\\nonumber
\end{eqnarray}
\begin{eqnarray}
%%%
\mathcal{M}_{{DP\to BV}}^{{P}} &=& \frac{g_{BDP}}{m_P}\bar{u}_3q_{\mu}u_1^{\mu}\frac{1}{q^2-m_{\mathbb{P}}^2}
      ig_{PPV}\epsilon_4^{\nu\dag}(q_{\nu}-p_{4\nu}),\\\nonumber\\
%%%
\mathcal{M}_{{DP\to BV}}^{{V}} &=& -i\frac{g_{BDV}}{m_V}\bar{u}_3\gamma^5(\gamma^{\nu}u_{1}^{\mu}-\gamma^{\mu}u_1^{\nu})q_{\mu}\nonumber\\
      &&\times\frac{g_{\nu\beta}-q_{\nu}q_{\beta}/m_{V}^2}{q^2-m_{V}^2}
      \frac{g_{VVP}}{m_V}\varepsilon^{\lambda\beta\alpha\delta}q_{\lambda}p_{4\alpha}\epsilon_{4\delta}^{\dag}.\\\nonumber
%%%
%\mathcal{M}_{\mathbb{DV\to BP}}^{\mathbb{P}} &=& \frac{g_{BDP}}{m_P}\bar{u}_3u_{1}^{\mu}q_{\mu}\frac{1}{q^2-m_{\mathbb{P}}^2}
%      ig_{PPV}(\epsilon_2^{\alpha}q_{\alpha}+p_{4\alpha}\epsilon_2^{\alpha}),\nonumber\\\\
%%%
%\mathcal{M}_{\mathbb{DV\to BP}}^{\mathbb{V}} &=& -\frac{g_{BDV}}{m_V}\bar{u}_3\gamma^5(\gamma^{\nu}u_{1}^{\mu}-\gamma^{\mu}u_1^{\nu})q_{\mu}\nonumber\\
%      &&\times\frac{g_{\nu\beta}-q_{\nu}q_{\beta}/m_{\mathbb{V}}^2}{q^2-m_{\mathbb{V}}^2}
%      ig_{VVP}\varepsilon^{\lambda\delta\alpha\beta}p_{2\lambda}\epsilon_{2\delta}q_{\alpha},\\
%%%
%\mathcal{M}_{\mathbb{DV\to BP}}^{\mathbb{B}} &=& \left\{g_{BBV}\bar{u}_4\gamma^{\alpha}\epsilon_{2\alpha}
%      +\frac{f_{BBV}}{4m^*}\bar{u}_4(\gamma^{\alpha}\gamma^{\beta}-\gamma^{\beta}\gamma^{\alpha})p_{2\alpha}\epsilon_{2\beta}\right\}\nonumber\\
%      &&\times\frac{-i}{\rlap\slash{q}-m_{\mathbb{B}}}\frac{g_{BDP}}{m_P}q_{\mu}u_1^{\mu},\\
%%%
\end{eqnarray}

\section{Numerical results}\label{sec3}

After prepared all the scattering amplitudes, we next adopt the wave functions obtained in Ref. \cite{Chen:2023qlx} to numerically calculate the partial decay widths of the predicted $P_{c\bar{s}}$ molecules. As is well known, the decay width can be very sensitive with the interactions and the wave functions of the initial and final states. Our results can be helpful to distinguish molecular states from compact pentaquarks due to their different wave functions. In addition, the probabilities for the $D$-wave components of all predicted molecular candidates are less than 1\% \cite{Chen:2023qlx}, it can hardly affect the decay behaviors, our results show that the difference of the total decay widths are less than 2\% whether considering the $D-$wave contributions or not, and the decay branching ratios for the discussed decay channels remain nearly unchanged. In the following, we only use the $S$-wave functions to present our results.

\begin{center}
% Requires \usepackage{graphicx}
  %\includegraphics[width=5.4in]{XisD1.eps}
\includegraphics[width=3.2in]{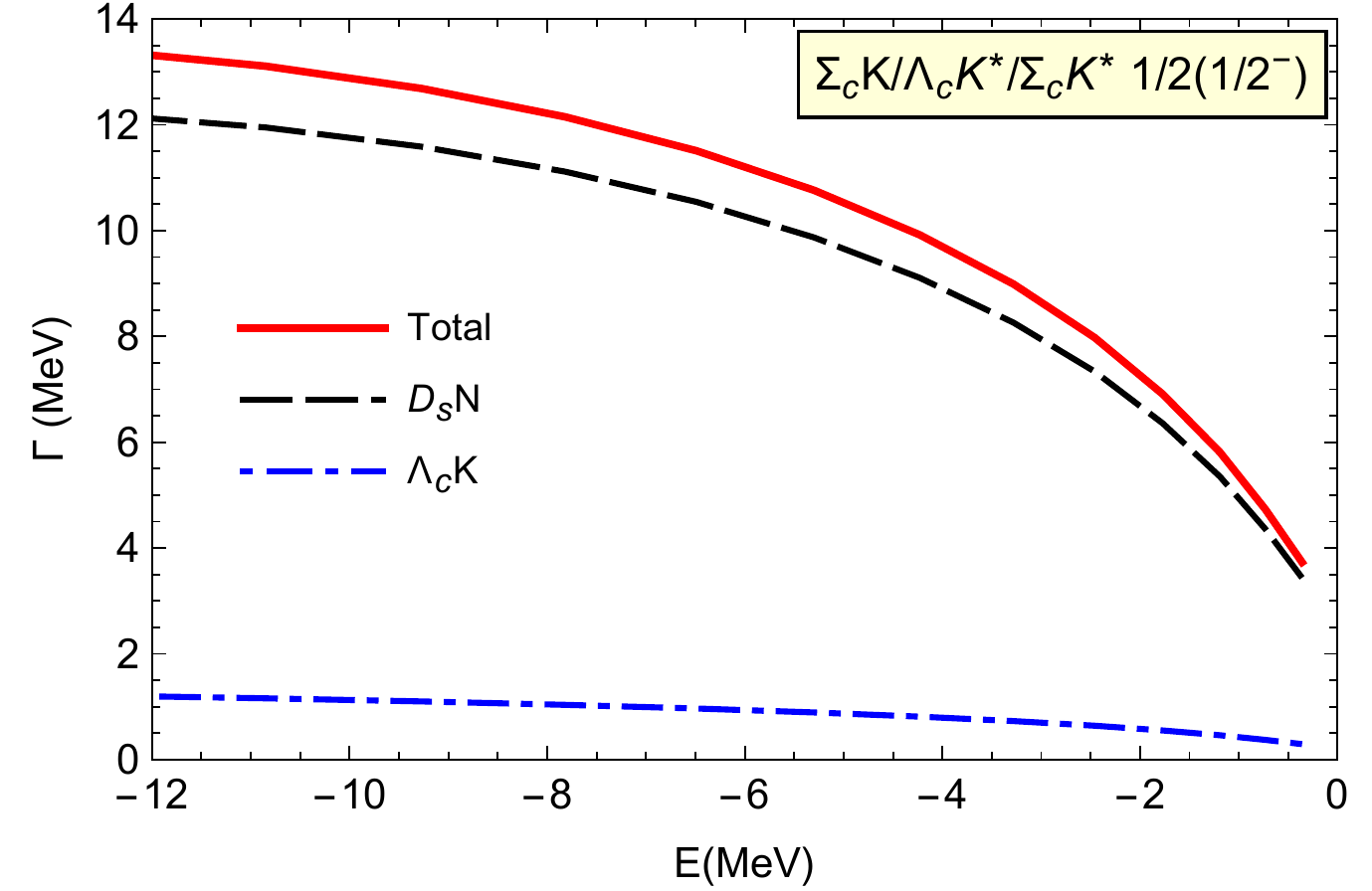}\\
\includegraphics[width=3.2in]{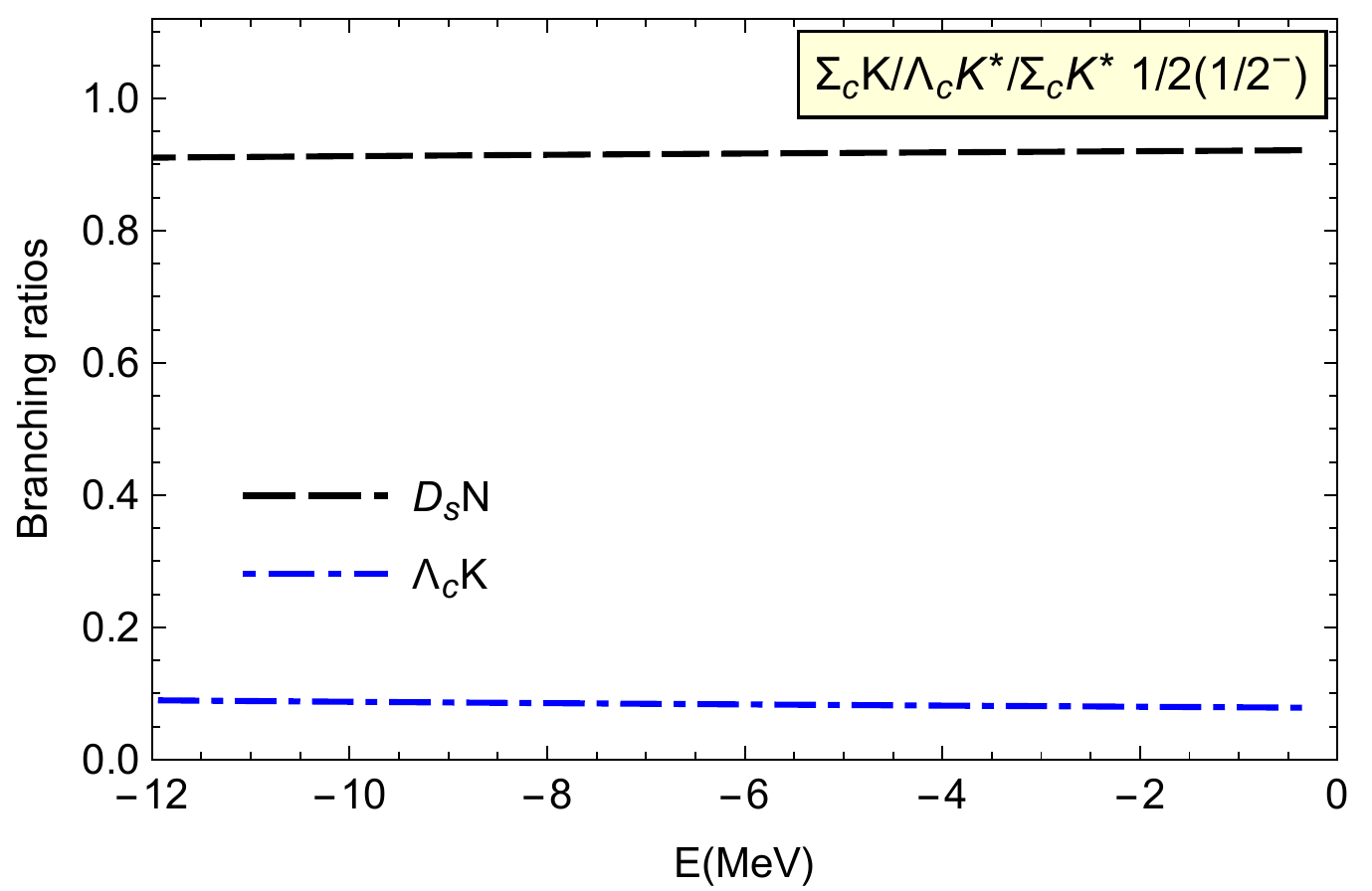}
\figcaption{\label{sigmak}The two-body strong decay width for the coupled $\Sigma_cK/\Lambda_cK^*/\Sigma_cK^*$ molecule with $I(J^P)=1/2(1/2^-)$.}
\end{center}

For the coupled $\Sigma_cK/\Lambda_cK^*/\Sigma_cK^*$ molecule with $I(J^P)=1/2(1/2^-)$, when the binding energy varies from 0 to $-12$ MeV, the probabilities for the $\Sigma_cK$ component is around 95\% or larger, therefore, the $\Sigma_cK$ component can be play an important role in the decay properties.

In Figure \ref{sigmak}, we present the two-body strong decay widths for the coupled $\Sigma_cK/\Lambda_cK^*/\Sigma_cK^*$ molecule with $I(J^P)=1/2(1/2^-)$. The $S-$wave two-body strong decay modes include $D_sN$ and $\Lambda_cK$. When the binding energy varies from 0 to $-12$ MeV, the total decay width is around ten MeV, and the $D_sN$ is the dominant decay channel.

Compared to the $D_s N$ mode, the partial decay width for the $\Lambda_c K$ final states is slightly smaller, being less than 2 MeV. The reasons are as follows: the light $\rho$ exchange interactions appear in the tensor terms $f_{\Lambda_c \Sigma_c \rho}/2m_B$, which are suppressed by the mass of the heavy baryon, $m_B$. The exchange interactions involving the heavy $\Xi_c^{(\prime)}$ baryons are also suppressed due to the large mass of the exchanged baryon.

In addition, we compare the decay widths in the single $\Sigma_cK$ channel, and we find the results are very similar, therefore, the coupled channel effects play a minor role for this coupled channel molecule.

\begin{center}
% Requires \usepackage{graphicx}
  %\includegraphics[width=5.4in]{XisD1.eps}
\includegraphics[width=3.2in]{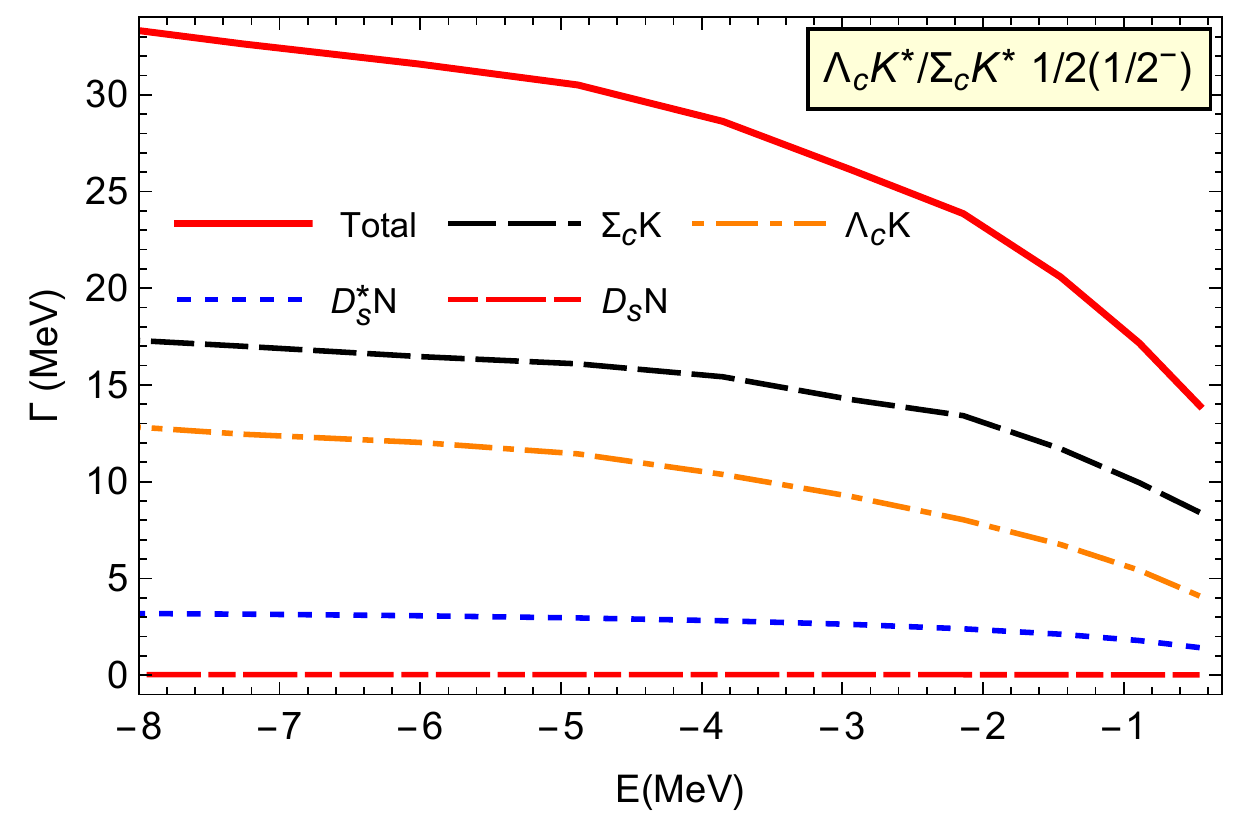}\\
\includegraphics[width=3.2in]{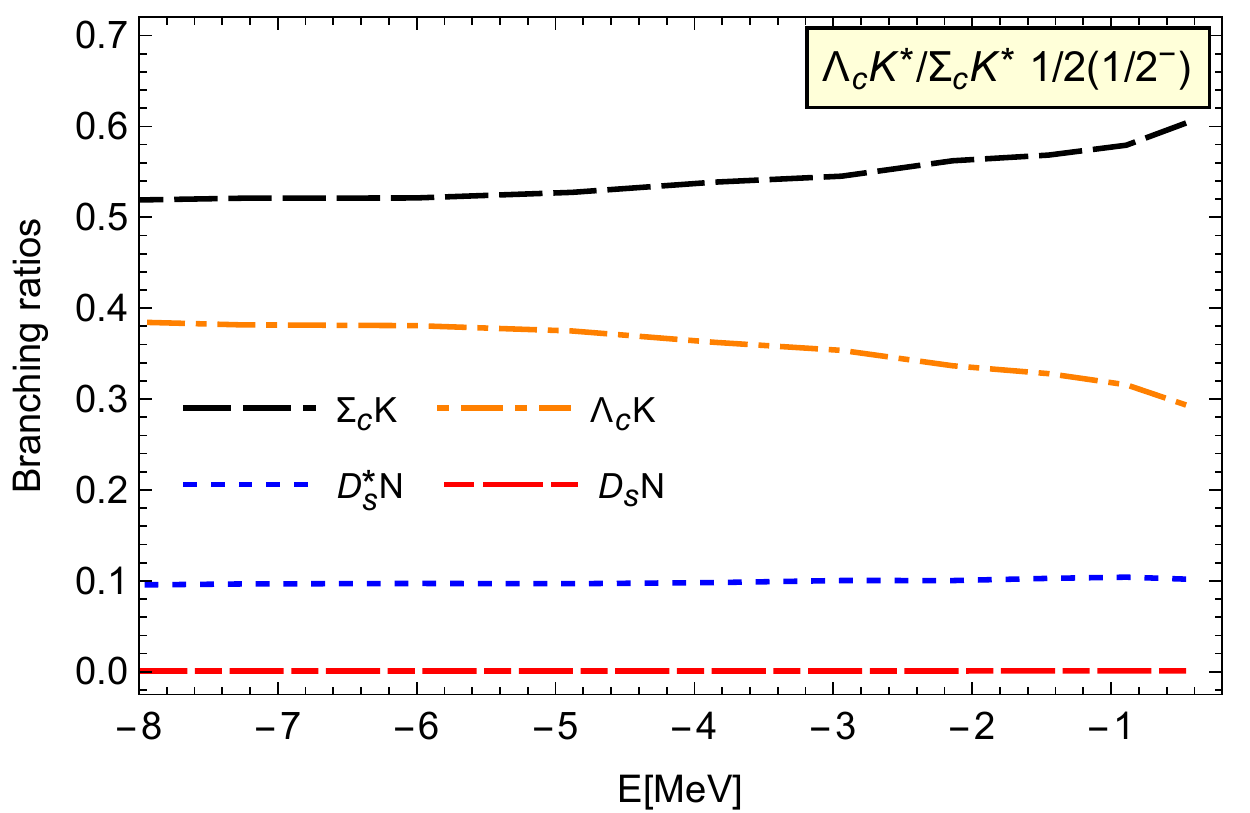}
\figcaption{\label{lamk1}The two-body strong decay width for the coupled $\Lambda_cK^*/\Sigma_cK^*$ molecule with $I(J^P)=1/2(1/2^-)$.}
\end{center}

In Figure \ref{lamk1}, we present the two-body strong decay widths for the coupled $\Lambda_cK^*/\Sigma_cK^*$ molecule with $I(J^P)=1/2(1/2^-)$. We consider four decay modes: the $D_sN$, $D_s^*N$, $\Sigma_cK$, and $\Lambda_cK$ channels. In the binding energy region of $E>-8$ MeV, the total decay width varies from 15 MeV to 30 MeV. Among these four decay channels, the $\Sigma_cK$ is the dominant decay channel, accounting for over 50\% of the total decay width. The large partial width arises from contributions due to light meson exchanges, particularly the $\pi$ exchange interactions in the $\Lambda_c K^* \to \Sigma_c K$ decay process. The secondary and tertiary dominant decay channels are $\Lambda_cK$ and $D_s^*N$ channels, respectively, with their ratios reaching around 40\% and 10\%. The decay width for the $D_sN$ is less than 1 MeV. Most importantly, the branching ratios for all the discussed channels are almost independent with the binding energy.

In addition, we find that the $S-$wave $\Sigma_cK^*$ component significantly contributes to the total decay width due to the important $\pi$ exchange interactions in the $\Sigma_cK^*\to\Lambda_cK$ decay process, even though its probability is less than 10\% in the binding energy range $E>-8$ MeV \cite{Chen:2023qlx}. The contributions from the $S-$wave $\Sigma_cK^*$ component reduce the total decay widths for this loosely bound molecular state as a partial coherence cancellation occurs between the interactions from the $\Lambda_cK^*$ and $\Sigma_cK^*$ components. From the above analysis, we can conclude that the coupled channel effects have a significant impact on the decay properties for this state.

\begin{center}
% Requires \usepackage{graphicx}
  %\includegraphics[width=5.4in]{XisD1.eps}
\includegraphics[width=3.2in]{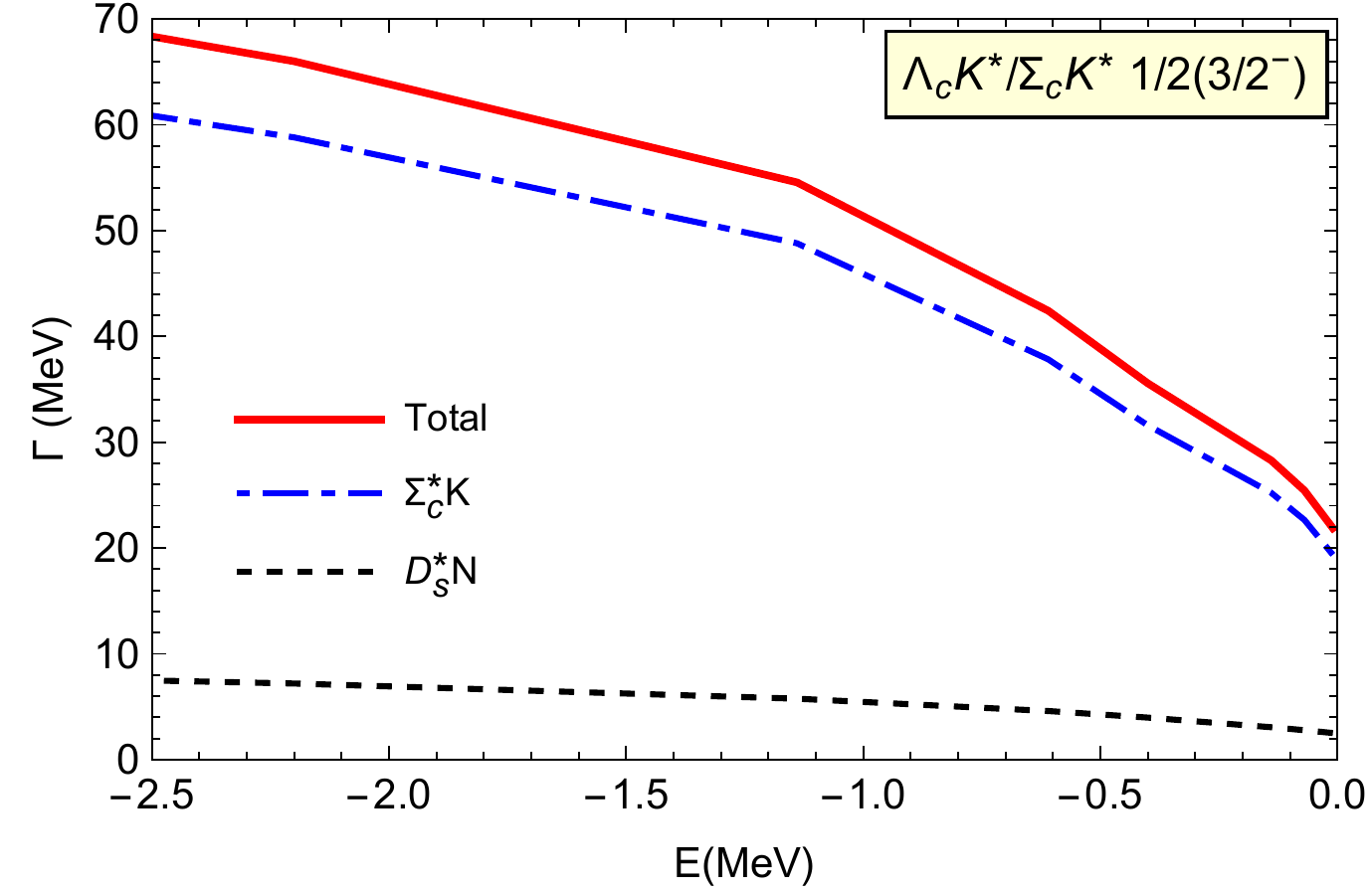}\\
\includegraphics[width=3.2in]{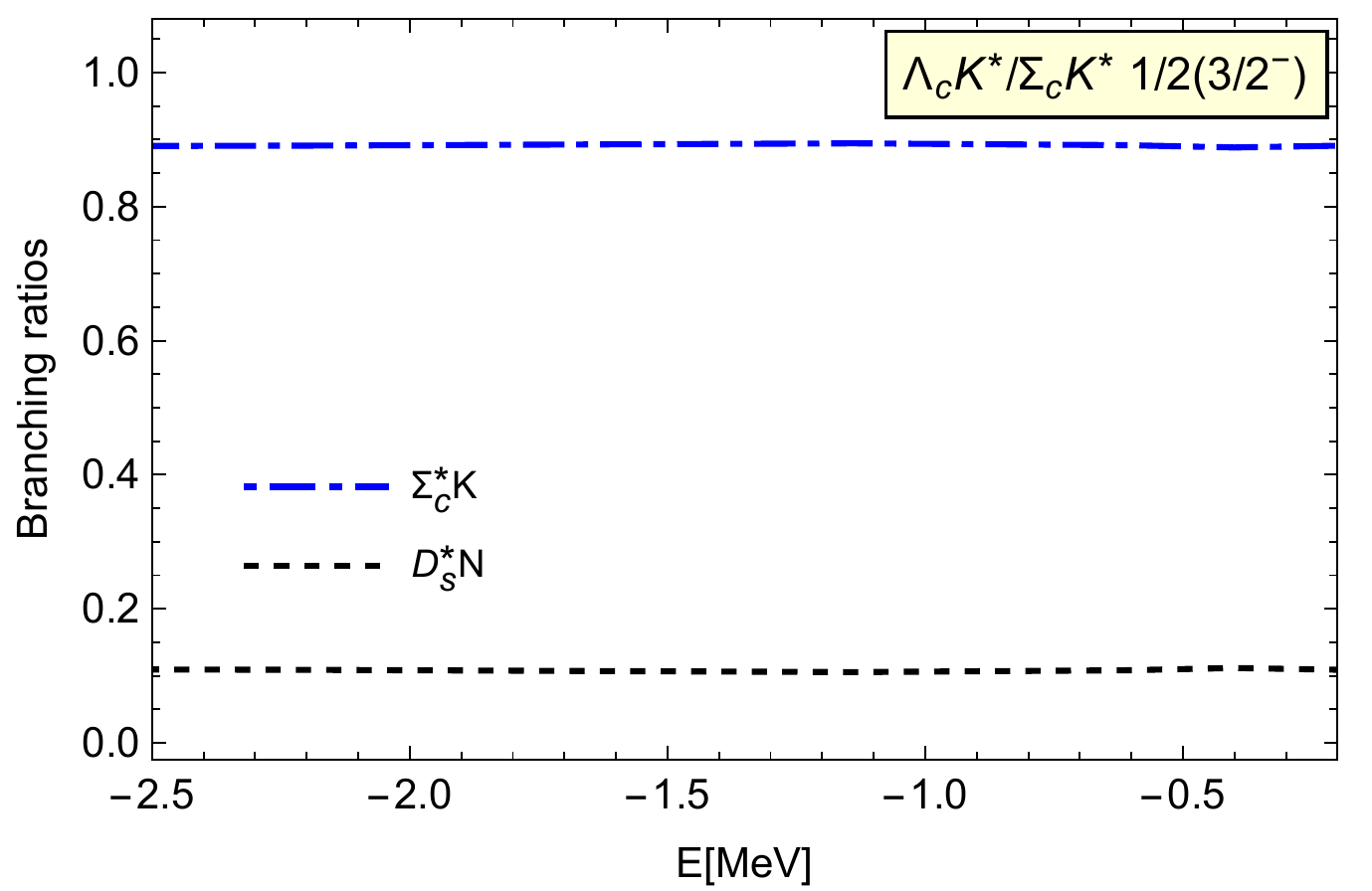}
\figcaption{\label{lamk2}The two-body strong decay width for the coupled $\Lambda_cK^*/\Sigma_cK^*$ molecule with $1/2(3/2^-)$.}
\end{center}

For the coupled $\Lambda_cK^*/\Sigma_cK^*$ molecule with $1/2(3/2^-)$, we consider the $\Sigma_c^*K$ and $D_s^*N$ decay modes. In Figure \ref{lamk2}, we present the decay width dependence of the binding energy. Here, the total decay width is around several tens MeV. The dominant decay channel is the $\Sigma_c^*K$, which contributes approximately 90\% of the total decay width.  The large branching ratio results from the significant $\pi$ exchange interaction, along with a coherent growth existing in the $\Lambda_c K^* \to \Sigma_c^* K$ and $\Sigma_c K^* \to \Sigma_c^* K$ interactions. Thus, the current results show the importance of the coupled channel effects again.

For the $D_s^*N$ channel, there don't exist the light mesons exchanges interactions as collected in Table \ref{channels}, the partial decay width is much smaller, it is around several MeV as shown in Figure \ref{lamk2}.

\begin{center}
% Requires \usepackage{graphicx}
  %\includegraphics[width=5.4in]{XisD1.eps}
\includegraphics[width=3.2in]{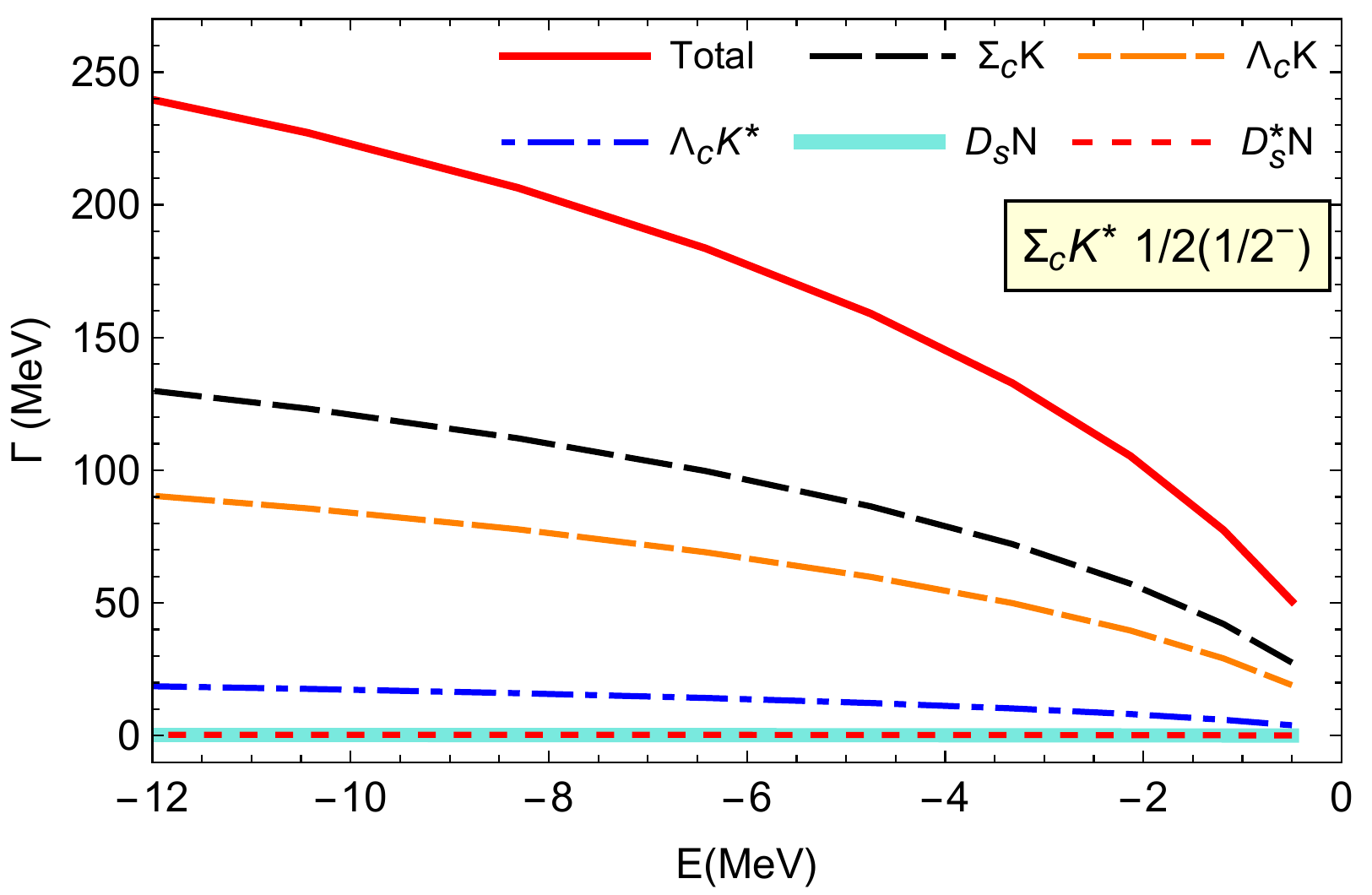}
\includegraphics[width=3.2in]{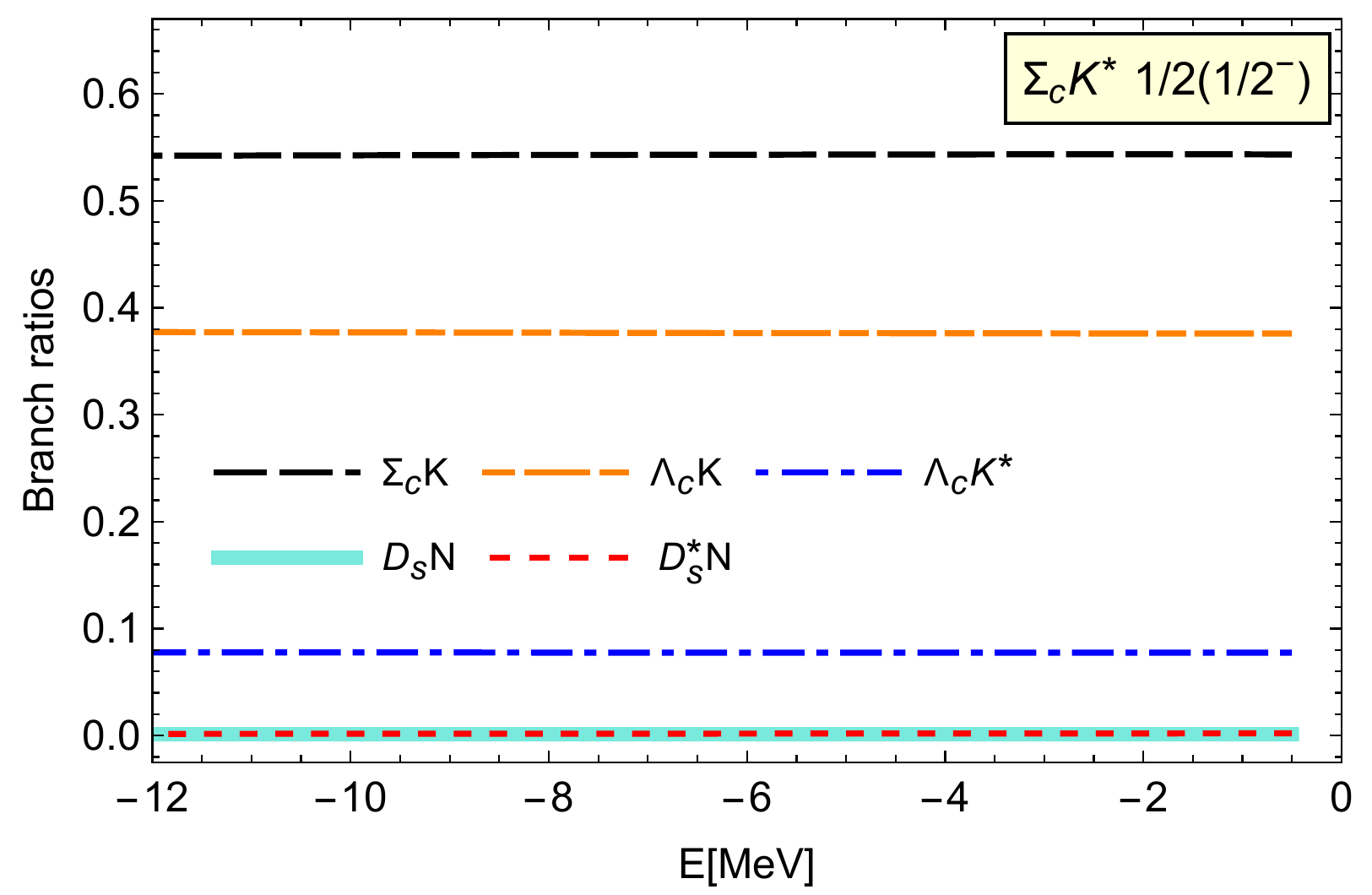}
\figcaption{\label{Sigmakx1}The two-body strong decay width for the single $\Sigma_cK^*$ molecules with $I(J^P)=1/2(1/2^-)$.}
\end{center}

In our previous work \cite{Chen:2023qlx}, we also predict three possible single $\Sigma_cK^*$ molecules with $I(J^P)=1/2(1/2^-)$, $1/2(3/2^-)$, and $3/2(1/2^-)$. In Fig. \ref{Sigmakx1}, we present their two-body strong decay widths for the $\Sigma_cK^*$ molecule with $I(J^P)=1/2(1/2^-)$, where the the binding energy is taking as $E>-12$ MeV. There are five decay channels via $S-$wave interactions, the $D_sN$, $D_s^*N$, $\Lambda_cK$, $\Lambda_cK^*$, and $\Sigma_cK$ channels. The total decay width can reach up to two hundred MeV within the binding energy $E>-12$ MeV. The $\Sigma_cK$ is the main decay channels, followed by the $\Lambda_cK$ and $\Lambda_cK^*$ channels. The corresponding branching ratios are 54\%, 38\%, and 8\%, respectively. In contrast, the decay widths for the $D_s^*N$ and $D_sN$ channels are very small, less than 1 MeV within the same binding energy range. We also find that the branching ratios for the discussed decay channels are not sensitive to the choice of the binding energy.

The above results can be explained as follows: as shown in Table \ref{channels}, the important interactions from the exchange of $\pi$ and $\rho$ mesons in the $\Sigma_cK^*\to\Sigma_c K(\Lambda_cK^{(*)})$ decay processes lead to the large decay widths for the $\Lambda_cK$, $\Lambda_cK^*$, and $\Sigma_cK$ decay modes. In addition, compared to the $\Sigma_cK$ and $\Lambda_cK$ channels, the phase space for the $\Sigma_cK^*\to\Lambda_cK^*$ decay process is much smaller, which is why the corresponding decay width is also much smaller.

\begin{center}
% Requires \usepackage{graphicx}
  %\includegraphics[width=5.4in]{XisD1.eps}
\includegraphics[width=3.2in]{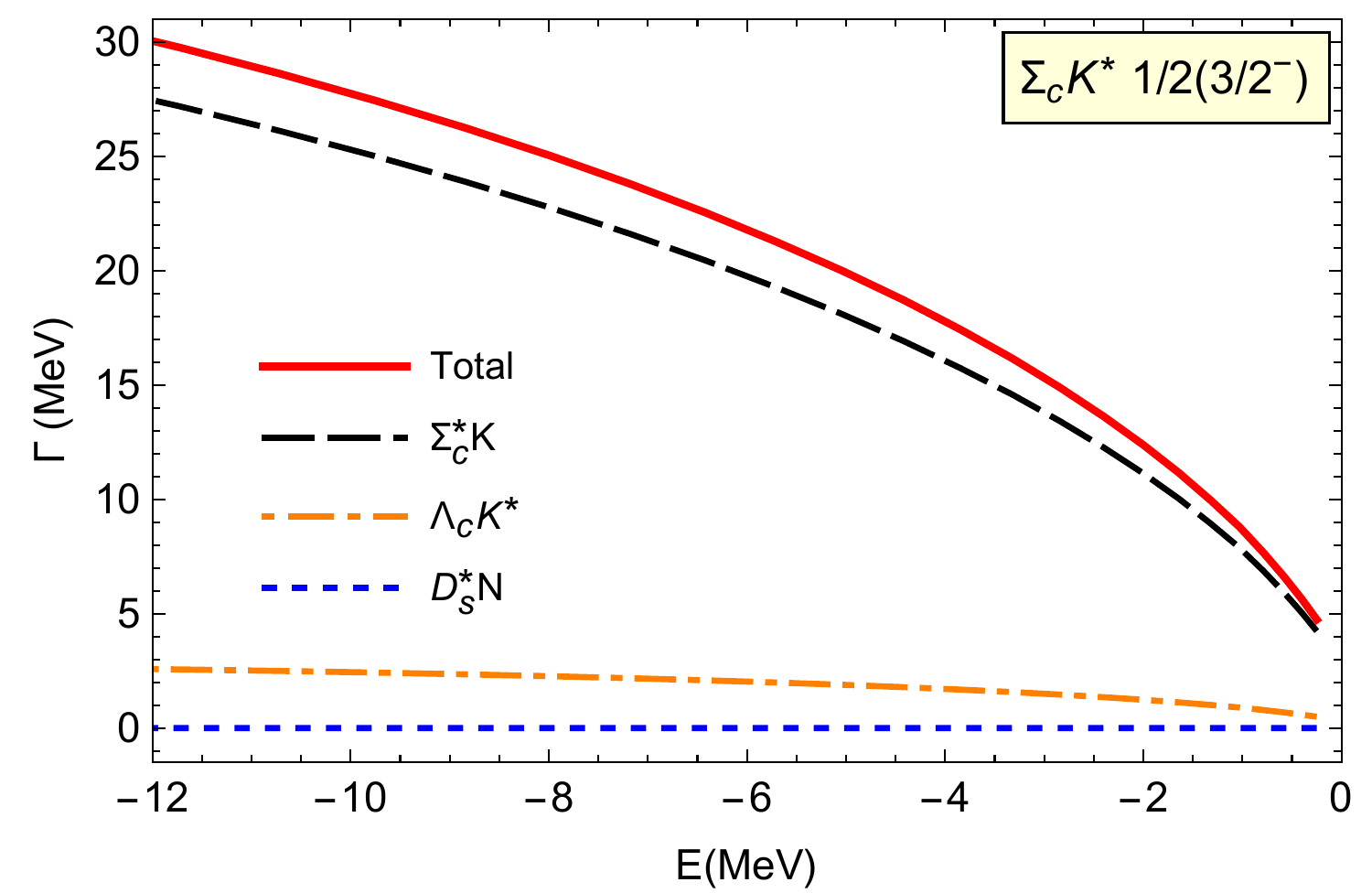}
\includegraphics[width=3.2in]{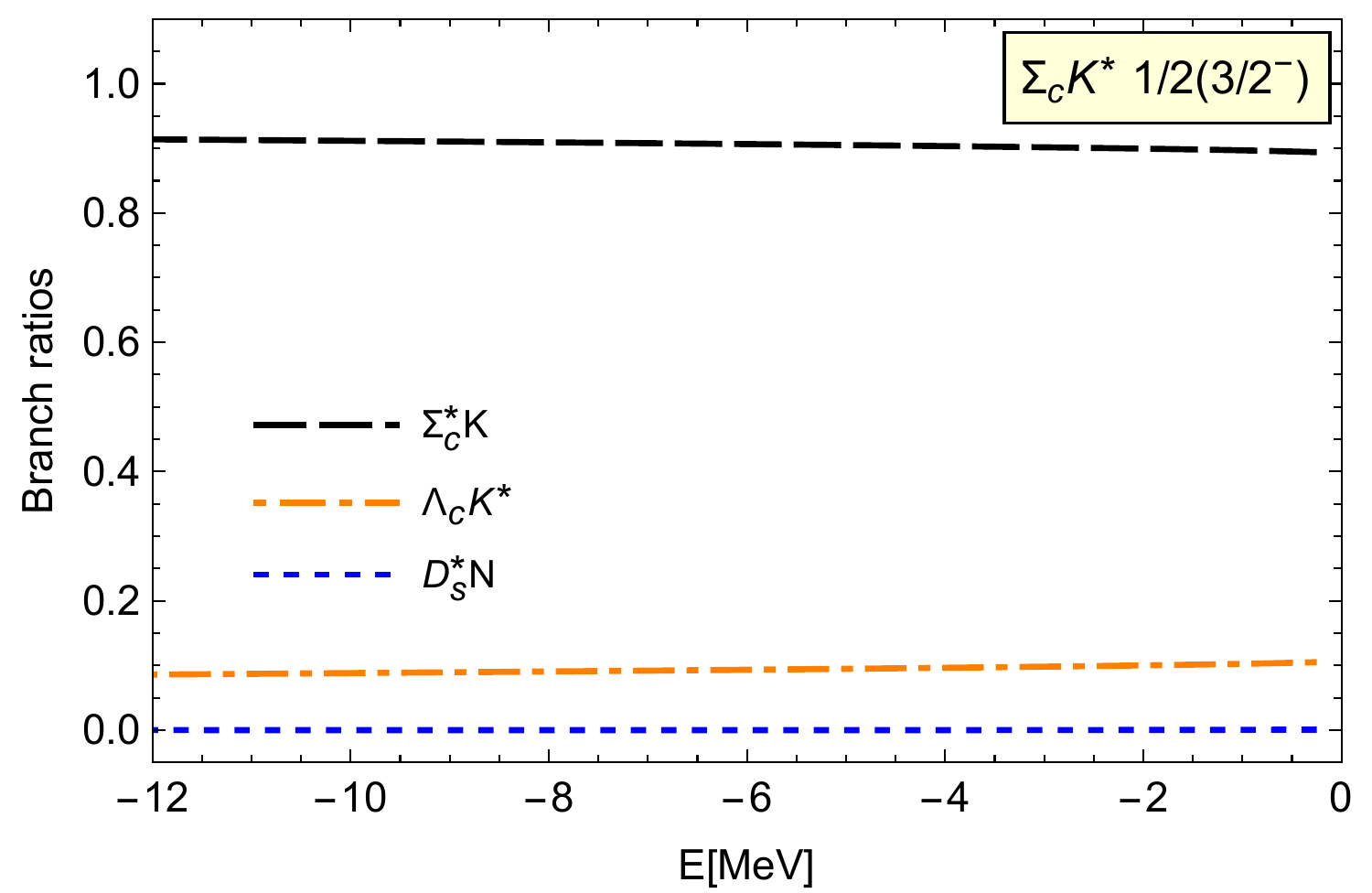}
\figcaption{\label{Sigmakx2}The two-body strong decay width for the single $\Sigma_cK^*$ molecules with $I(J^P) =1/2(3/2^-)$.}
\end{center}

For the $\Sigma_cK^*$ molecule with $1/2(3/2^-)$, we discuss the $D_s^*N$, $\Lambda_cK^*$, and $\Sigma_c^*K$ decay modes. Our results indicate that the total decay width is around several tens MeV when the binding energy is taken as $E > -12$ MeV, as shown in Figure \ref{Sigmakx2}. This is much smaller than the $\Sigma_cK^*$ molecule with $1/2(1/2^-)$, and the main reason is that the $\Sigma_cK^*$ molecule with $1/2(3/2^-)$ decays into the $\Sigma_cK$ and $\Lambda_cK$ channels through $D-$wave interactions. The $\Sigma_c^*K$ is the dominant decay mode, which occupies around 90\% of the total decay width, due to the important $\pi$ exchange interactions. For the $\Lambda_cK^*$ decay mode, the partial decay width is on the order of several MeV, even though light meson exchange interactions are involved. One important reason for this small decay width is that the phase space for this decay channel is much smaller than for the $\Sigma_c^*K$ mode. In addition, there is the strong coherent cancellation between the meson exchange and baryon exchange interactions in the $\Sigma_cK^*$ molecule with $1/2(3/2^-)$ decaying into the $D_s^*N$ final states, which results in a decay width less than 0.01 MeV when the binding energy is $E > -12$ MeV. In addition, in Ref. \cite{Liu:2022qzc}, authors adopted the resonance group method in the quark delocalization color screening model and predicted the masses and the widths for the possible $\Sigma_cK^*$ bound states with $I(J^P)=1/2(1/2^-)$ and $1/2(3/2^-)$. They also found that the decay width for the higher spin is much smaller.

\begin{center}
% Requires \usepackage{graphicx}
  %\includegraphics[width=5.4in]{XisD1.eps}
\includegraphics[width=3.2in]{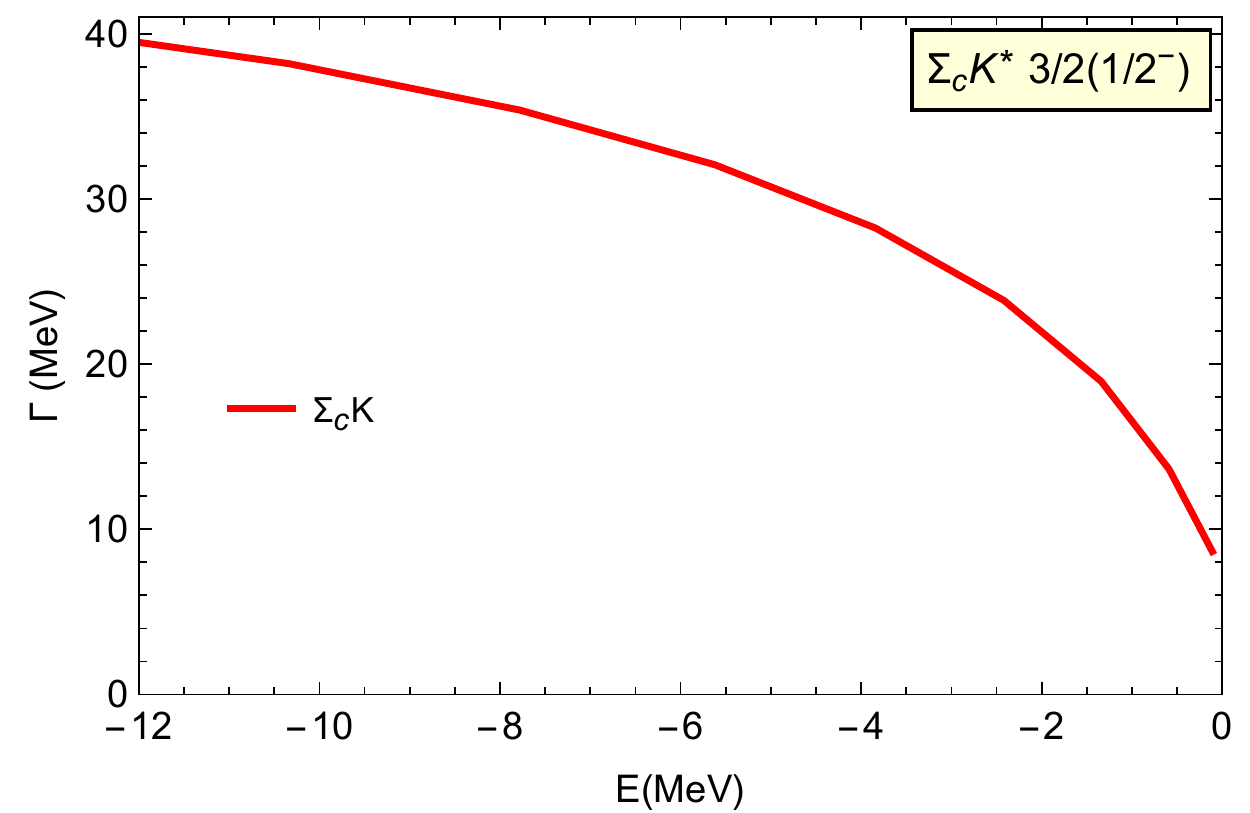}
\figcaption{\label{Sigmakx3}The two-body strong decay width for the single $\Sigma_cK^*$ molecules with $I(J^P)=3/2(1/2^-)$.}
\end{center}

For the $\Sigma_cK^*$ molecule with $3/2(1/2^-)$, the $\Sigma_cK$ is the only two-body strong decay mode via $S-$wave interaction. The decay width is around several tens MeV with $E>-12$ MeV as shown in Figure \ref{Sigmakx3}. Compared to the $\Sigma_cK^*$ molecular with $1/2(1/2^-)$, the $\pi$ and $\rho$ exchange interactions here are about twice as weak, as shown in Table \ref{channels}, which leads to the much smaller decay width for the $\Sigma_c K^*$ molecule with $I(J^P) = 3/2(1/2^-)$.

\section{Discussion and conclusion}\label{sec4}

Studying the strong decay properties of molecular states is crucial for understanding their internal structure. Different decay channels and branching ratios provide clues about the internal structure and properties of the hadronic molecular states.

In this work, we employ the effective Lagrangian approach to study the two-body strong decay behaviors of the predicted open-charm molecules $Y_c K^{(*)}$ \cite{Chen:2023qlx}. And we also consider the coupled channel effects. The corresponding numerical results are summarized in Table \ref{summary}. We find the decay width for the coupled $\Sigma_cK/\Lambda_cK^*/\Sigma_cK^*$ molecule with $1/2(1/2^-)$ is around ten MeV, with the $D_s N$ channel being dominant. The decay widths for the coupled $\Lambda_cK^*/\Sigma_cK^*$ molecules with $1/2(1/2^-)$ and $1/2(3/2^-)$ are on the order of several tens of MeV, with the dominant channels being $\Sigma_c K$ and $\Sigma_c^* K$, respectively. For the same binding energies, the coupled $\Lambda_cK^*/\Sigma_cK^*$ molecules with $1/2(1/2^-)$ exhibits a narrower decay width, because the coupled-channel effects play an important role, here. The decay widths for the $\Sigma_cK^*$ molecules with $1/2(1/2^-)$, $1/2(3/2^-)$, and $3/2(1/2^-)$ are on the order of one hundred MeV and a few tens of MeV, respectively. The dominant decay channels are $\Sigma_c^{(*)}(\Lambda_c) K$, driven by light meson exchanges such as $\pi$ and $\rho$. Additionally, the branching ratios for all the discussed channels are not sensitive to the binding energy.

\end{multicols}
\begin{center}
\tabcaption{\label{summary} A summary of the two-body decay properties for the possible $Y_cK^{(*)}$ molecules. Here, $E$ and $\Gamma$ stands for the range of the binding energy and the total width, respectively. The units of the binding energy $E$ and the decay width $\Gamma$ are MeV.}
\footnotesize
\renewcommand{\arraystretch}{1.9}
\begin{tabular*}{177mm}{@{\extracolsep{\fill}}lc|c|c|llll}
\toprule[1pt]\toprule[1pt]
States &$I(J^P)$   &$E$    &$\Gamma$   &\multicolumn{4}{c}{Channels (Branch ratios)}    \\\hline
$\Sigma_cK/\Lambda_cK^*/\Sigma_cK^*$   &$1/2(1/2^-)$  &$(0,-12)$    &$(4,13)$
   &$D_sN(91\%)$  &$\Lambda_cK(9\%)$  \\
$\Lambda_cK^*/\Sigma_cK^*$             &$1/2(1/2^-)$    &$(0,-8)$   &$(14,33)$
 &$\Sigma_cK(54\%)$     &$\Lambda_cK(36\%)$     &$D_s^*N(10\%)$      &$D_sN(<1\%)$\\
                                       &$1/2(3/2^-)$   &$(0,-2.5)$   &$(25,70)$    &$\Sigma_c^*K(90\%)$ &$D_s^*N$ (10\%)\\
$\Sigma_cK^*$                           &$1/2(1/2^-)$  &$(0,-12)$    &$(50,245)$
  &$\Sigma_cK(54\%)$   &$\Lambda_cK(38\%)$    &$\Lambda_cK^*(8\%)$     &$D_s^{(*)}N(<1\%)$  \\
                                       &$1/2(3/2^-)$  &$(0,-12)$    &$(5,30)$
   &$\Sigma_c^*K(91\%)$   &$\Lambda_cK^*(9\%)$    &$D_s^*N(<1\%)$\\
                                       &$3/2(1/2^-)$   &$(0,-12)$   &$(10,39)$     &$\Sigma_cK(100\%)$\\
\bottomrule[1pt]\bottomrule[1pt]
\end{tabular*}%
\end{center}

\begin{multicols}{2}

In this work, we only focus on the two-body strong decay properties for the $Y_cK^{(*)}$ molecules, in fact, the three-body decay modes via $K^*\to K\pi$ may be important for the molecules involving the $K^*$ components. Compared to the two-body decay modes, the phase space could be smaller than the two-body decays, which can lead to smaller decay widths. We would like to calculate the three-body decay widths in our next work.

In addition, we need to mention that the effective Lagrangians are taken the same forms of those constructed in the well-established $SU(3)$ symmetry. We apply the $SU(4)$ symmetry to relate the coupling constants. As shown in Eqs. (\ref{lag1})-(\ref{lag2}), we use the masses of pseudoscalar and vector mesons instead of $m_{\pi}$, $m_{\rho}$, and $m_{\omega}$ in the original $SU(3)$ expressions, this could be regarded as a correction to the strongly broken $SU(4)$ flavor symmetry \cite{ Shen:2019evi,Yalikun:2021dpk,Wang:2022oof,Yue:2024paz}. Here, we also discuss the uncertainties of the coupling constants as the $SU(4)$ symmetry breaking. In order to estimate the effects of the $SU(4)$ symmetry breaking, we compare the coupling constant adopted in the correction SU(4) symmetry relations and the values extracted from the experimental widths, such as the decay widths for the $\Sigma_c^{(*)}\to \Lambda_c\pi$ processes. And we find the uncertainties of the coupling constants are less than 20\%. With this reason, we then choose three groups of typical values to present our results, where the coupling constant $g_H$ related to the heavy quarks are taken as $0.8g_H$, $g_H$, and $1.2g_H$, respectively, as shown in Table \ref{num1}. Here, we can find the obtained widths have the same order of magnitudes, and the branch ratios for all the discussed modes are almost the same.

In summary, the information obtained can not only provide insights into the interactions between the components, but also serve as a valuable guide for experimental searches. We expect the experiments to verify our predictions, processes such as $B\to\Lambda_c(\Sigma_c)\bar{\Lambda}_cK$, $\Lambda_b \to\Lambda_cK\bar{K}\pi$ have the potential to search for these possible $Y_cK^{(*)}$ molecules.

\begin{center}
\tabcaption{\label{num1} A summary of the total decay widths for the possible $Y_cK^{(*)}$ molecules via two-body strong decay interactions. Here, the units of the binding energy $E$ and the decay width $\Gamma$ are MeV.}
\footnotesize
\renewcommand{\arraystretch}{1.7}
\begin{tabular*}{86mm}{@{\extracolsep{\fill}}lc|cccc}
\toprule[1pt]\toprule[1pt]
States &$I(J^P)$   &$E$   &$\Gamma(0.8g_H)$  &$\Gamma(g_H)$ &$\Gamma(1.2g_H)$      \\\hline
$\Sigma_cK/\Lambda_cK^*/\Sigma_cK^*$   &$1/2(1/2^-)$  &$-5.30$ &4.51 &$10.76$    &22.10 \\
$\Lambda_cK^*/\Sigma_cK^*$             &$1/2(1/2^-)$  &-4.88   &21.32   &$30.51$     &40.82 \\
                                       &$1/2(3/2^-)$  &-1.14  &35.26  &$54.56$    &77.91   \\
$\Sigma_cK^*$                           &$1/2(1/2^-)$ &-4.75   &102.72  &$159.02$     &227.02 \\
                                       &$1/2(3/2^-)$  &-5.71 &13.39    &$21.27$    &32.10 \\
                                       &$3/2(1/2^-)$  &-5.62  &21.13 &$32.07$   &44.86   \\
\bottomrule[1pt]\bottomrule[1pt]
\end{tabular*}
\end{center}

\end{multicols}

\vspace{10mm}

\begin{multicols}{2}

\end{multicols}

\clearpage

\end{CJK*}
\end{document}